\documentclass[aps,prc,reprint,showpacs,floatfix,superscriptaddress]{revtex4-1}
\usepackage{amsmath}
\usepackage{amssymb}
\usepackage{graphicx}
\usepackage{hyperref}

\hypersetup{
    colorlinks=true,
    linkcolor=blue,
    citecolor=red,
    filecolor=magenta,      
    urlcolor=cyan,
}

\usepackage{breakurl}
\usepackage{lipsum}
\DeclareGraphicsExtensions{ .eps, .eps.gz,
                            .ps, .ps.gz, .png }

\begin{document}
\verb| |\\  
\title{Extracting the spin polarizabilities of the proton by measurement of Compton double-polarization observables}
\author{D.~Paudyal}
\affiliation{University of Regina, Regina, Saskatchewan S4S 0A2, Canada}
\author{P.P.~Martel\thanks{}}
\email{martel@uni-mainz.de}
\affiliation{Mount Allison University, Sackville, New Brunswick E4L 1E6, Canada}
\affiliation{Institut f{\"u}r Kernphysik, University of Mainz, D-55099 Mainz, Germany}
\author{G.M.~Huber}
\affiliation{University of Regina, Regina, Saskatchewan S4S 0A2, Canada}
\author{D.~Hornidge}
\affiliation{Mount Allison University, Sackville, New Brunswick E4L 1E6, Canada}
\author{S.~Abt}
\affiliation{Institut f{\"u}r Physik, University of Basel, CH-4056, Basel, Switzerland}
\author{P.~Achenbach}
\affiliation{Institut f{\"u}r Kernphysik, University of Mainz, D-55099 Mainz, Germany}
\author{P.~Adlarson}
\affiliation{Institut f{\"u}r Kernphysik, University of Mainz, D-55099 Mainz, Germany}
\author{F.~Afzal}
\affiliation{Helmholtz-Institut f{\"u}r Strahlen- und Kernphysik, University of Bonn, D-53115 Bonn, Germany}
\author{Z.~Ahmed}
\affiliation{University of Regina, Regina, Saskatchewan S4S 0A2, Canada}
\author{C.S.~Akondi}
\affiliation{Kent State University, Kent, Ohio 44242-0001, USA}
\author{J.R.M.~Annand }
\affiliation{SUPA School of Physics and Astronomy, University of Glasgow, Glasgow G12 8QQ, UK}
\author{H.J.~Arends}
\affiliation{Institut f{\"u}r Kernphysik, University of Mainz, D-55099 Mainz, Germany}
\author{M.~Bashkanov}
\affiliation{Department of Physics, University of York, Heslington, York, Y010 5DD, UK}
\author{R.~Beck}
\affiliation{Helmholtz-Institut f{\"u}r Strahlen- und Kernphysik, University of Bonn, D-53115 Bonn, Germany}
\author{M.~Biroth}
\affiliation{Institut f{\"u}r Kernphysik, University of Mainz, D-55099 Mainz, Germany}
\author{N.S.~Borisov}
\affiliation{Joint Institute for Nuclear Research, 141980 Dubna, Russia}
\author{A.~Braghieri }
\affiliation{INFN Sezione di Pavia, I-27100 Pavia, Italy}
\author{W.J.~Briscoe}
\affiliation{The George Washington University, Washington, DC 20052-0001, USA}
\author{F.~Cividini}
\affiliation{Institut f{\"u}r Kernphysik, University of Mainz, D-55099 Mainz, Germany}
\author{S.~Costanza }
\affiliation{INFN Sezione di Pavia, I-27100 Pavia, Italy}
\author{C.~Collicott }
\affiliation{Dalhousie University, Halifax, Nova Scotia B3H 4R2, Canada}
\affiliation{Saint Mary's University, Halifax, Nova Scotia B3H 3C3, Canada}
\author{A.~Denig}
\affiliation{Institut f{\"u}r Kernphysik, University of Mainz, D-55099 Mainz, Germany}
\author{M.~Dieterle}
\affiliation{Institut f{\"u}r Physik, University of Basel, CH-4056, Basel, Switzerland}
\author{E.J.~Downie}
\affiliation{The George Washington University, Washington, DC 20052-0001, USA}
\author{P.~Drexler}
\affiliation{Institut f{\"u}r Kernphysik, University of Mainz, D-55099 Mainz, Germany}
\author{M.I.~Ferretti-Bondy}
\affiliation{Institut f{\"u}r Kernphysik, University of Mainz, D-55099 Mainz, Germany}
\author{S.~Gardner}
\affiliation{SUPA School of Physics and Astronomy, University of Glasgow, Glasgow G12 8QQ, UK}
\author{S.~Garni}
\affiliation{Institut f{\"u}r Physik, University of Basel, CH-4056, Basel, Switzerland}
\author{D.I.~Glazier}
\affiliation{SUPA School of Physics and Astronomy, University of Glasgow, Glasgow G12 8QQ, UK}
\author{D.~Glowa}
\affiliation{SUPA School of Physics and Astronomy, University of Edinburgh, Edinburgh EH9 3FD, UK}
\author{I.~Gorodnov}
\affiliation{Joint Institute for Nuclear Research, 141980 Dubna, Russia}
\author{W.~Gradl}
\affiliation{Institut f{\"u}r Kernphysik, University of Mainz, D-55099 Mainz, Germany}
\author{S.~G{\"u}nther}
\affiliation{Institut f{\"u}r Physik, University of Basel, CH-4056, Basel, Switzerland}
\author{G.M.~Gurevich}
\affiliation{Institute for Nuclear Research, 125047 Moscow, Russia}
\author{D.~Hamilton}
\affiliation{SUPA School of Physics and Astronomy, University of Glasgow, Glasgow G12 8QQ, UK}
\author{L.~Heijkenskj{\"o}ld}
\affiliation{Institut f{\"u}r Kernphysik, University of Mainz, D-55099 Mainz, Germany}
\author{A.~K{\"a}ser}
\affiliation{Institut f{\"u}r Physik, University of Basel, CH-4056, Basel, Switzerland}
\author{V.L.~Kashevarov}
\affiliation{Institut f{\"u}r Kernphysik, University of Mainz, D-55099 Mainz, Germany}
\affiliation{Joint Institute for Nuclear Research, 141980 Dubna, Russia}
\author{S.~Kay}
\affiliation{University of Regina, Regina, Saskatchewan S4S 0A2, Canada}
\author{I.~Keshelashvili}
\affiliation{Institut f{\"u}r Physik, University of Basel, CH-4056, Basel, Switzerland}
\author{R.~Kondratiev}
\affiliation{Institute for Nuclear Research, 125047 Moscow, Russia}
\author{M.~Korolija}
\affiliation{Rudjer Boskovic Institute, HR-10000 Zagreb, Croatia}
\author{B.~Krusche}
\affiliation{Institut f{\"u}r Physik, University of Basel, CH-4056, Basel, Switzerland}
\author{A.B.~Lazarev }
\affiliation{Joint Institute for Nuclear Research, 141980 Dubna, Russia}
\author{J.M.~Linturi}
\affiliation{Institut f{\"u}r Kernphysik, University of Mainz, D-55099 Mainz, Germany}
\author{V.~Lisin}
\affiliation{Institute for Nuclear Research, 125047 Moscow, Russia}
\author{K.~Livingston}
\affiliation{SUPA School of Physics and Astronomy, University of Glasgow, Glasgow G12 8QQ, UK}
\author{S.~Lutterer}
\affiliation{Institut f{\"u}r Physik, University of Basel, CH-4056, Basel, Switzerland}
\author{I.J.D.~MacGregor}
\affiliation{SUPA School of Physics and Astronomy, University of Glasgow, Glasgow G12 8QQ, UK}
\author{R.~Macrae}
\affiliation{SUPA School of Physics and Astronomy, University of Glasgow, Glasgow G12 8QQ, UK}
\author{J.~Mancell}
\affiliation{SUPA School of Physics and Astronomy, University of Glasgow, Glasgow G12 8QQ, UK}
\author{D.M.~Manley}
\affiliation{Kent State University, Kent, Ohio 44242-0001, USA}
\author{V.~Metag}
\affiliation{II. Physikalisches Institut, University of Giessen, D-35392 Giessen, Germany}
\author{W.~Meyer}
\affiliation{Institut f{\"u}r Experimentalphysik, Ruhr Universit{\"a}t, 44780 Bochum, Germany}
\author{R.~Miskimen}
\affiliation{University of Massachusetts Amherst, Amherst, Massachusetts 01003, USA}
\author{E.~Mornacchi}
\affiliation{Institut f{\"u}r Kernphysik, University of Mainz, D-55099 Mainz, Germany}
\author{C.~Mullen}
\affiliation{SUPA School of Physics and Astronomy, University of Glasgow, Glasgow G12 8QQ, UK}
\author{A.~Mushkarenkov}
\affiliation{University of Massachusetts Amherst, Amherst, Massachusetts 01003, USA}
\affiliation{Institute for Nuclear Research, 125047 Moscow, Russia}
\author{A.B.~Neganov }
\affiliation{Joint Institute for Nuclear Research, 141980 Dubna, Russia}
\author{A.~Neiser}
\affiliation{Institut f{\"u}r Kernphysik, University of Mainz, D-55099 Mainz, Germany}
\author{M.~Oberle}
\affiliation{Institut f{\"u}r Physik, University of Basel, CH-4056, Basel, Switzerland}
\author{M.~Ostrick}
\affiliation{Institut f{\"u}r Kernphysik, University of Mainz, D-55099 Mainz, Germany}
\author{P.B.~Otte}
\affiliation{Institut f{\"u}r Kernphysik, University of Mainz, D-55099 Mainz, Germany}
\author{P.~Pedroni }
\affiliation{INFN Sezione di Pavia, I-27100 Pavia, Italy}
\author{A.~Polonski}
\affiliation{Institute for Nuclear Research, 125047 Moscow, Russia}
\author{A.~Powell}
\affiliation{SUPA School of Physics and Astronomy, University of Glasgow, Glasgow G12 8QQ, UK}
\author{S.N.~Prakhov}
\affiliation{Institut f{\"u}r Kernphysik, University of Mainz, D-55099 Mainz, Germany}
\affiliation{University of California Los Angeles, Los Angeles, California 90095-1547, USA}
\author{A.~Rajabi}
\affiliation{University of Massachusetts Amherst, Amherst, Massachusetts 01003, USA}
\author{G.~Reicherz}
\affiliation{Institut f{\"u}r Experimentalphysik, Ruhr Universit{\"a}t, 44780 Bochum, Germany}
\author{G.~Ron}
\affiliation{Racah Institute of Physics, Hebrew University of Jerusalem, Jerusalem 91904, Israel}
\author{T.~Rostomyan}
\affiliation{Institut f{\"u}r Physik, University of Basel, CH-4056, Basel, Switzerland}
\author{A.~Sarty}
\affiliation{Saint Mary's University, Halifax, Nova Scotia B3H 3C3, Canada}
\author{C.~Sfienti}
\affiliation{Institut f{\"u}r Kernphysik, University of Mainz, D-55099 Mainz, Germany}
\author{M.H.~Sikora}
\affiliation{SUPA School of Physics and Astronomy, University of Edinburgh, Edinburgh EH9 3FD, UK}
\author{V.~Sokhoyan}
\affiliation{Institut f{\"u}r Kernphysik, University of Mainz, D-55099 Mainz, Germany}
\affiliation{The George Washington University, Washington, DC 20052-0001, USA}
\author{K. ~Spieker}
\affiliation{Helmholtz-Institut f{\"u}r Strahlen- und Kernphysik, University of Bonn, D-53115 Bonn, Germany}
\author{O.~Steffen}
\affiliation{Institut f{\"u}r Kernphysik, University of Mainz, D-55099 Mainz, Germany}
\author{I.I.~Strakovsky}
\affiliation{The George Washington University, Washington, DC 20052-0001, USA}
\author{Th.~Strub}
\affiliation{Institut f{\"u}r Physik, University of Basel, CH-4056, Basel, Switzerland}
\author{I.~Supek}
\affiliation{Rudjer Boskovic Institute, HR-10000 Zagreb, Croatia}
\author{A.~Thiel}
\affiliation{Helmholtz-Institut f{\"u}r Strahlen- und Kernphysik, University of Bonn, D-53115 Bonn, Germany}
\author{M.~Thiel}
\affiliation{Institut f{\"u}r Kernphysik, University of Mainz, D-55099 Mainz, Germany}
\author{A.~Thomas}
\affiliation{Institut f{\"u}r Kernphysik, University of Mainz, D-55099 Mainz, Germany}
\author{M.~Unverzagt}
\affiliation{Institut f{\"u}r Kernphysik, University of Mainz, D-55099 Mainz, Germany}
\author{Yu.A.~Usov}
\affiliation{Joint Institute for Nuclear Research, 141980 Dubna, Russia}
\author{S.~Wagner}
\affiliation{Institut f{\"u}r Kernphysik, University of Mainz, D-55099 Mainz, Germany}
\author{N.K.~Walford}
\affiliation{Institut f{\"u}r Physik, University of Basel, CH-4056, Basel, Switzerland}
\author{D.P.~Watts}
\affiliation{Department of Physics, University of York, Heslington, York, Y010 5DD, UK}
\author{D.~Werthm{\"u}ller}
\affiliation{Department of Physics, University of York, Heslington, York, Y010 5DD, UK}
\author{J.~Wettig}
\affiliation{Institut f{\"u}r Kernphysik, University of Mainz, D-55099 Mainz, Germany}
\author{L.~Witthauer}
\affiliation{Institut f{\"u}r Physik, University of Basel, CH-4056, Basel, Switzerland}
\author{M.~Wolfes}
\affiliation{Institut f{\"u}r Kernphysik, University of Mainz, D-55099 Mainz, Germany}
\author{L.~Zana}
\affiliation{Thomas Jefferson National Accelerator Facility, Newport News, Virginia 23606, USA}
\collaboration{A2 Collaboration}
\noaffiliation
\date{\today}
\setlength{\abovedisplayskip}{0cm}

\begin{abstract}
  The Compton double-polarization observable $\Sigma_{2z}$ has been measured for
  the first time in the $\Delta(1232)$ resonance region using a circularly
  polarized photon beam incident on a longitudinally polarized proton target at
  the Mainz Microtron. This paper reports these results, together with the
  model-dependent extraction of four proton spin polarizabilities from fits to
  additional asymmetry data using either a dispersion relation calculation or a
  baryon chiral perturbation theory calculations, with the weighted average of
  these two fits resulting in:
  $\gamma_{E1E1} = -2.87 \pm 0.52$, $\gamma_{M1M1} = 2.70 \pm 0.43$,
  $\gamma_{E1M2} = -0.85 \pm 0.72$ and $\gamma_{M1E2} = 2.04 \pm 0.43$,
  in units of $10^{-4}~\mathrm{fm}^{4}$. 
\end{abstract}

\pacs{25.20.Lj, 13.40.-f, 13.60.Fz, 13.88.+e}

\maketitle

The electromagnetic interaction of a photon with a nucleon can be studied
through Compton scattering experiments. It is best described using
an effective Hamiltonian expanded in terms of the incident photon energy.
Structure observables of these composite systems are experimentally
accessible by elastically scattering real photons from the nucleon in Real
Compton Scattering (RCS). Over decades, RCS has been established as a
benchmark for understanding the ground-state properties of the nucleon, such
as the magnetic moment. However, the leading-order properties that are
sensitive to the internal quark dynamics of the nucleon are still
poorly understood experimentally. This paper uses RCS in the
$\Delta(1232)$ resonance region as a probe to understand some
internal structure observables of a nucleon, the nucleon polarizabilities.
These are fundamental properties that describe how its internal structure
deforms under an applied electromagnetic field~\cite{pasq07,hagel16}.

The electromagnetic field of the photon induces transitions of certain
definite multipolarities while attempting to deform the nucleon. The effective
Hamiltonian at second order in incident photon energy, $E_{\gamma}$, depends on
the electric and magnetic scalar polarizabilities, $\alpha_{E1}$ and
$\beta_{M1}$, and at third order depends on the spin polarizabilities (SPs).

The third-order effective Hamiltonian term in the spin-dependent interaction is
\begin{multline} 
{H_{eff}^{(3)}} = -4\pi \Big[{\frac {1}{2}}\gamma_{E1E1}
	\vec{\sigma \cdot}(\vec{E}\times\dot{\vec{E}})+\frac{1}{2}
	\gamma_{M1M1}\vec{\sigma \cdot}(\vec{H}\times\dot{\vec{H}}) \\ 
 - \gamma_{M1E2}E_{ij}\sigma_{i}H_{j}
	+\gamma_{E1M2}H_{ij}\sigma_{i}E_{j} \Big],
\end{multline}\label{eq:c2_13}
where $\dot{\vec E} $, $\dot{\vec H}$, $E_{ij}$ and $H_{ij}$ are the
partial derivatives with respect to time and space defined as
$\dot{\vec E} = \partial_{t} \vec E$, $\dot{\vec H} = \partial_{t} \vec H$,
$ E_{ij} = \frac{1}{2} (\partial_{i} E_{j} + \partial_{j} E_{i})$ and
$ H_{ij} = \frac{1}{2} (\partial_{i} H_{j} + \partial_{j} H_{i})$, and
$\gamma_{E1E1}$, $\gamma_{M1M1}$, $\gamma_{M1E2}$ and $\gamma_{E1M2}$ are the
four SPs. The physics behind these leading-order SPs involves the excitation of
the spin-$\frac {1}{2}$ target nucleon to some intermediate state ($\Delta $ or
$N^{\star}$) via an electric or magnetic ($E1$ or $M1$) dipole transition and a
successive de-excitation back to a spin-$\frac {1}{2} $ nucleon final state
via an electric or magnetic dipole ($E1$ or $M1$) or quadrupole ($E2$ or $M2$)
transition. These internal structure constants are manifestations of the spin
structure of the nucleon, which parameterize the ``stiffness'' of the
nucleon's spin against the electromagnetically induced deformations relative
to the spin axis.

Measurements of two linear combinations of these four SPs---the forward spin
polarizability, $\gamma_{0}$~\cite{ahrens01,dutz03}, and the backward spin
polarizability, $\gamma_{\pi}$~\cite{camen02}---have been reported for the
proton by several experiments. An extraction of the individual proton SPs was
recently published via measurement of the double-polarization Compton
asymmetry---$\Sigma_{2x}$---using a transversely polarized proton target at the
Mainz Microtron (MAMI)~\cite{martel15} in conjunction with the $\gamma_{0}$ and
$\gamma_{\pi}$ results and measurement of the beam-polarization Compton
asymmetry $\Sigma_{3}$ performed at the LEGS facility~\cite{blanpied01}.
This paper describes an improvement to the extraction of these proton SPs from
the measurement of the double-polarization asymmetry $\Sigma_{2z}$ using a
longitudinally polarized proton target at MAMI. $\Sigma_{2z}$ is defined as
\begin{equation}
{\Sigma}_{2z} = \frac{1}{ P^{\gamma}_{circ}\cdot P^{t}_{z}}
	\left[\frac {(N^{R}_{+z} +N^{L}_{-z}) - (N^{L}_{+z} +N^{R}_{-z})}
	{(N^{R}_{+z} +N^{L}_{-z}) + (N^{L}_{+z} + N^{R}_{-z})}\right], 
\end{equation}
where $N^{R}_{\pm z}$ and $N^{L}_{\pm z}$ are the normalized yield for
right-handed and left-handed helicity states of the beam with the target
polarized in the $\pm z$ direction, and $P^{\gamma}_{circ} $ and $P^{t}_{z}$ are
the degrees of the photon beam circular polarization and target polarization,
respectively. 

The experiment was performed in the A2 hall at MAMI~\cite{kaiser08,janko06}, a
facility composed of a cascade of three Race Track Microtrons that can provide
both unpolarized and longitudinally polarized electron beams with energies up
to $1.6~$GeV~\cite{kaiser08}. The longitudinally polarized electron beam was
produced by irradiating a strained GaAsP III-V semiconductor with circularly
polarized laser light~\cite{aulen97}. A $180^{\circ}$ polarization flip was
provided by reversing the helicity of the laser light with a Pockels cell at a
rate of approximately 1~Hz. A standard Mott polarimeter~\cite{tiouk11},
installed near the MAMI accelerator cascade, was used for polarization
measurements. The average beam polarization was
$86.8\pm 0.1 \%$~\cite{paudyal17}. For this measurement, a $450~$MeV polarized
electron beam passed through an alloy radiator of cobalt and iron, producing
circularly polarized Bremsstrahlung photons. The photon polarization,
$P_{\gamma}$, was determined by the helicity transfer relationship
\begin{equation}
  P_{\gamma}=P_{e}\frac{4E_{\gamma}E_{e}-E_{\gamma}^{2}}
  {4E_{e}^{2}-4E_{\gamma}E_{e}+3E_{\gamma}^{2}},
\end{equation}
where $P_{e}$ is the electron beam polarization, $E_{e}$ is the electron beam
energy, and $E_{\gamma}$ is the energy of the radiated photon. $E_{\gamma}$
was determined by detecting the Bremsstrahlung electrons in the tagged photon
spectrometer~\cite{mcg08}, and only photons in the energy range
$E_{\gamma} = 265 - 305~$MeV were used for this analysis. The previously
mentioned flip of the electron beam polarization direction results in a flip
of the photon beam helicity, which, given the relatively fast rate of 1~Hz,
provides the $\Sigma_{2z}$ asymmetry relatively free of systematic effects.
The photon beam was passed through a 2.5-mm-diameter lead collimator, resulting
in a beam spot size of 9~mm on the longitudinally polarized Frozen Spin Target
(FST)~\cite{thomas11} located in the center of the Crystal Ball spectrometer
(CB)~\cite{star01}. 

The FST used dynamic nuclear polarization, and its polarization was measured
with a nuclear magnetic resonance coil; both are described in detail in
Ref.~\cite{thomas11}. Polarization of up to 80\% and relaxation times of
nearly 1000 hours were achieved~\cite{tarbert08,mcg08}, and the direction of
proton polarization was reversed approximately once per week. While flipping
the photon helicity is enough to produce the $\Sigma_{2z}$ asymmetry,
additionally reversing the target polarization is useful to further remove
systematic effects. Polarization measurements were completed at the start
and end of each data taking period for different polarization orientations.
Corrections to the target polarization due to ice buildup on the NMR
coil~\cite{afzal19} were determined with $\pi^{0}$ asymmetries as well as
comparisons of unpolarized and polarized total inclusive and $\pi^{0}$ cross
sections~\cite{tech19}. To reflect inconsistencies between these methods, a
liberal systematic error of 10\% for the target polarization was utilized.

Data were collected during two beamtimes in 2014 and 2015 using the nearly
$4\pi$ CB-TAPS detector system~\cite{unverz09}: the CB as a central
calorimeter, and TAPS as a forward calorimeter. The CB consists of 672
optically isolated NaI(Tl) crystals with a truncated triangular pyramid shape
arranged in two hemispheres.  It covers about $94\%$ of $4\pi$ steradians and
an angular range of $21^{\circ} \leq \theta \leq 159^{\circ}$~\cite{blanpied01}.
TAPS consists of 366 hexagonal BaF$_{2}$ crystals and two inner rings totaling
72 PbWO$_{4}$ crystals and covers an angular range of
$2^{\circ} \leq \theta \leq 20^{\circ}$~\cite{nov91}. Charged particles were
identified using energy deposition in the particle identification detector and
tracked by a pair of multi-wire proportional chambers or TAPS-veto detectors
and their corresponding calorimetric detector. Although the CB-TAPS system
covers the angular range of $2-159^{\circ}$, there are regions near
the entrance and exit through the detectors that are less efficient. These
regions are: (i) the forward hole in the TAPS detector, $2-6^{\circ}$, and
(ii) the backward hole in the CB, $150-159^{\circ}$. Fiducial cuts were
applied to remove all the data from these angular regions of reduced detection
efficiency.

The Compton scattering channel, $\gamma p \rightarrow \gamma p$, has a simple
final state, but it is very important to correctly identify background from
competing reactions because its cross section is only about $1\%$ of the cross
section for the dominant $\pi^{0}$ photoproduction process. In addition, under
certain conditions, $\pi^{0}$ photoproduction can mimic the Compton scattering
signature if one of the $\pi^{0}$ decay photons escapes the detector, or if the
electromagnetic showers from the two photons overlap due to finite angular
resolution. The Compton channel was identified by selecting events having a
total deposited energy above $40~$MeV, where a single neutral and a single
charged track are reconstructed, with the former in coincidence with a hit in
the tagger. In order to remove uncorrelated events between CB-TAPS and the
photon tagger, the timing differences between the neutral track and hits in the
tagger were checked against a $20~$ns wide prompt (coincidence) window and a
$910~$ns wide random window split in two with one on either side of the prompt
peak. The random sample was normalized by the relative window widths and
subtracted from the prompt timing signal.

To eliminate competing backgrounds from coherent and incoherent Compton
scattering and $\pi^{0}$ photoproduction off of non-hydrogen nuclei in the FST
from the windows and shells of the cryostat material (mainly
$^{3}$He/$^{4}$He, $^{12}$C and $^{16}$O), separate data were taken by
inserting a carbon foam target with density $0.55~$g/cm$^{3}$ into the same
cryostat and the normalized yield was subtracted. A base scaling factor was
determined by the ratio of live-time corrected tagger scalers for the butanol
and carbon data sets. Comparison of $\pi^{0}$ photoproduction simulations with
the data showed that a correction to this ratio of approximately 10\% was
necessary to account for a higher contribution from the helium in the target to
this background. The carbon target density was chosen such that the number of
nucleons equals the number of non-hydrogen nucleons from the $^{3}$He/$^{4}$He,
$^{12}$C and $^{16}$O in the target. To remove background from $\pi^{0}$
photoproduction off of the proton, the coincidence of a recoil charged track
in addition to the neutral track was required, as mentioned above.
However, since protons suffer a significant amount of energy loss when they
travel from the event vertex through the target material, a $^{3}$He/$^{4}$He
refrigeration bath, various cryostat shells and a longitudinal holding coil
on their way to a detector crystal, the analysis was limited to an incident
photon energy range of $E_{\gamma} = 265 - 305~$MeV. Further details
on the background cuts, subtractions, and normalization factors can be found
in Ref.~\cite{paudyal17,tech19}.
\begin{figure}[ht]
\centering
\includegraphics[width=8.6cm]{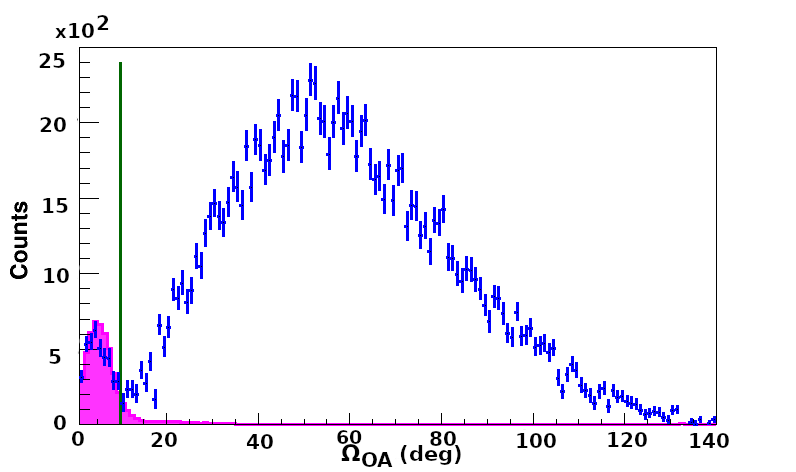}
\caption{Opening angle distribution for simulated Compton
scattering events (magenta) compared with the carbon-subtracted data (blue) at
$E_{\gamma} = 285 - 305~$MeV and over all Compton angles. A cut of
$10^{\circ}$ on the opening angle is indicated by the vertical line (green).}
\label{fig-oa}
\end{figure}

To identify events of interest, four-momentum conservation was used to
constrain the observed reaction kinematics. As the background varies
significantly across both energy and angle, their dependencies were studied.
The tagged photon energy bins below $\gamma p \rightarrow \pi^{0} \pi^{0} p$
threshold were divided into five $\theta$ bins, and were analyzed separately.
The opening angle ($\Omega_{OA}$), defined as the angle between the detected
proton, $\vec{p}_{recoil}$, and where the proton was expected assuming RCS
kinematics, $\vec{p}_{miss} = \vec{p}_{\gamma_i} - \vec{p}_{\gamma_f}$,
$\cos(\Omega_{OA})=\frac{\vec{p}_{miss}.\vec{p}_{recoil}}{\vec{|p}_{miss}|\times|\vec{p}_{recoil}|}$,
was used for a two-body reaction selection. The Monte Carlo simulated opening
angle results show a sharp peak around $5^{\circ}$, which is in good agreement
with the data. The large background, as seen in Fig.~\ref{fig-oa}, is mainly
due to the $\pi^{0}$ photoproduction process from the proton. This
can be suppressed by applying a $10^{\circ}$ opening angle cut, as indicated
by the green vertical line. 
\begin{figure}[ht]
\centering
\includegraphics[width=8.6 cm]{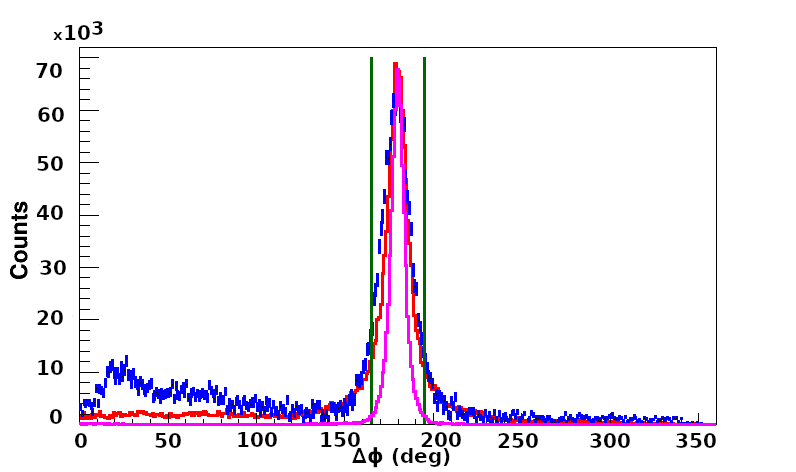}
\caption{Coplanarity distribution for simulated Compton scattering events
(magenta), and simulated $\pi^{0}$ events that were analyzed as if they were
a Compton photon (red), compared with the carbon-subtracted data (blue) at
$E_{\gamma} = 285 - 305~$MeV and over all Compton angles ($\Omega_{OA}$ cut from
Fig.~\ref{fig-oa} is applied).}
\label{fig-cop}
\end{figure}
The Compton coplanarity angle, defined as the difference in the azimuthal angles
of a scattered photon and a recoil proton,
$\Delta\phi =|\phi_{\gamma f}-\phi_{p}|$, was used to suppress additional
background. A cut on the fixed coplanarity angle,
$\Delta\phi = 180 \pm 15 ^{\circ}$, as indicated by the two vertical green lines
in Fig.~\ref{fig-cop}, was applied to the reconstructed events. For those events
with a single neutral and a single charged track, the missing mass is calculated
with
\begin{equation} {M_{miss}}^{2} = {\left(E_{\gamma_i}+m_{p} c^{2}-E_{\gamma_f
}\right)^{2}-\left(\vec{p}_{\gamma_i}-\vec{p}_{\gamma_f }\right)^{2}c^{2}},
\end{equation}
where ($E_{\gamma_i}, \vec{p}_{\gamma_i}c$) and ($E_{\gamma_f}, \vec{p}_{\gamma_f}c$)
are the four vectors of the incident and scattered photon, respectively, and
$m_{p}$ is the proton mass. 

\begin{figure}[ht]
\centering
\includegraphics[width=8.6 cm]{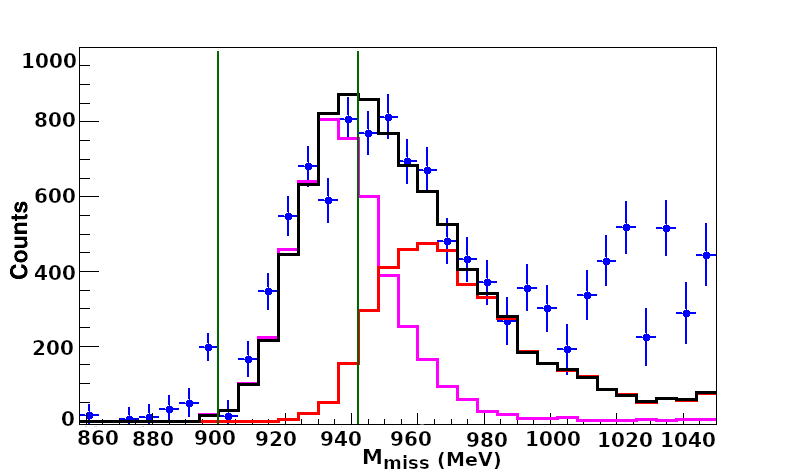}
\caption{Missing mass spectrum for carbon-subtracted data (blue), Monte Carlo
simulated results from Compton scattering (magenta) and $\pi^{0}$ photoproduction
(red) satisfying Compton cuts, and the sum of the two simulated contributions to
show an expected distribution (black), all at $\theta_{\gamma}=125-140^{\circ}$
and $E_{\gamma} = 285-305~$MeV. Two vertical lines (green) represent the missing
mass integration limit.}
\label{fig-mm}
\end{figure}
The carbon-subtracted $M_{miss}$ spectrum using the corrected carbon target
scaling factors~\cite{paudyal17,tech19} is shown in Fig.~\ref{fig-mm}.
Simulations of both Compton scattering and $\pi^{0}$ production were passed
through the same analysis chain, with the same cuts applied. The distributions
from these two reactions were added together (according to their known cross
section at a given energy and angle). From these spectra, there is clearly good
agreement of the data with the expected distribution up to $M_{miss}$ $\approx$
980~MeV/c$^{2}$. It is observed that $\pi^{0}$ photoproduction is the major
source of background above a $M_{miss}$ of approximately 940~MeV/c$^{2}$, and
hence it is necessary to set a clear upper $M_{miss}$ limit that coincides with
the turn-on point of this background.

Though $M_{miss}$ spectra can be integrated up to the most conservative limit of
938~MeV/c$^{2}$ (proton mass), the following steps were taken to maximize the
integrated yield. First, the lower $M_{miss}$ limit was fixed at 900~MeV/c$^{2}$
for each energy and angle bin. Second, the asymmetry determined using the
conservative $M_{miss}$ upper limit of 938~MeV/c$^{2}$ was taken as a reference.
Finally, the asymmetry was allowed to vary a maximum of 5$\%$ by moving the
$M_{miss}$ upper limit to higher values compared to the reference. This $\pm 5\%$
is based on the systematic uncertainties from the choice of carbon target length
and the ratio of $\pi^{0}$ photoproduction background to Compton scattering
determined from simulation. As the asymmetry shifts either up or down for
different bins, there is no concern about introducing a systematic shift from
the `correct' asymmetry. As an additional check the central value on the spin
polarizabilities, as extracted by the method described below, were compared
between the reference and final asymmetries, which indicated only small effects
on $\gamma_{E1E1}$ and $\gamma_{M1E2}$ of approximately 20\% of their errors and
negligible effects on the other two. The resulting final $M_{miss}$ upper limits
are between $940-948~$MeV/c$^{2}$, and further details on this work can be found
in Ref.~\cite{paudyal17,tech19}.

\begin{table}[ht]
\centering
\begin{tabular}{l|l|lll}
  \hline
  $E_{\gamma}$ (MeV)  & $\theta_{\gamma}$  &  $\Sigma_{2z}$  &  Rand.  &  Syst.  \\ \hline
                      & $ 87.5^{\circ}$    &  0.193          &  $\pm 0.056$  &  $\pm 0.024$  \\
                      & $102.5^{\circ}$    &  0.290          &  $\pm 0.040$  &  $\pm 0.035$  \\
  265--285            & $117.5^{\circ}$    &  0.402          &  $\pm 0.037$  &  $\pm 0.048$  \\
                      & $132.5^{\circ}$    &  0.672          &  $\pm 0.036$  &  $\pm 0.077$  \\ 
                      & $147.5^{\circ}$    &  0.672          &  $\pm 0.042$  &  $\pm 0.081$  \\ \hline
                      & $ 87.5^{\circ}$    &  0.121          &  $\pm 0.040$  &  $\pm 0.016$  \\
                      & $102.5^{\circ}$    &  0.279          &  $\pm 0.034$  &  $\pm 0.033$  \\
  285--305            & $117.5^{\circ}$    &  0.428          &  $\pm 0.038$  &  $\pm 0.048$  \\
                      & $132.5^{\circ}$    &  0.591          &  $\pm 0.029$  &  $\pm 0.066$  \\ 
                      & $147.5^{\circ}$    &  0.751          &  $\pm 0.046$  &  $\pm 0.085$  \\ \hline
\end{tabular}
\caption{Summary of results and uncertainties for the Compton $\Sigma_{2z}$ asymmetry.}
\label{table:Asy}
\end{table}

\begin{figure}[ht]
\centering
\includegraphics[width=8.6 cm]{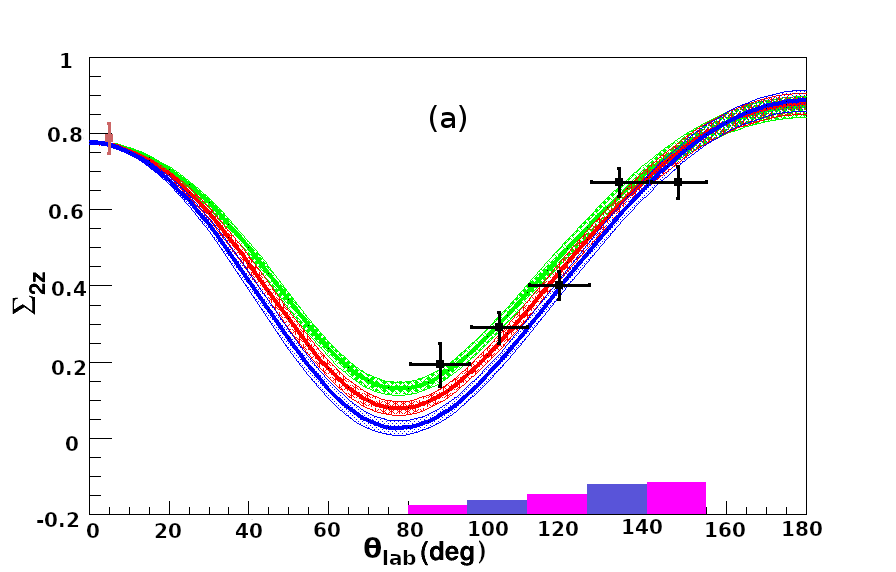}\\
\includegraphics[width=8.6 cm]{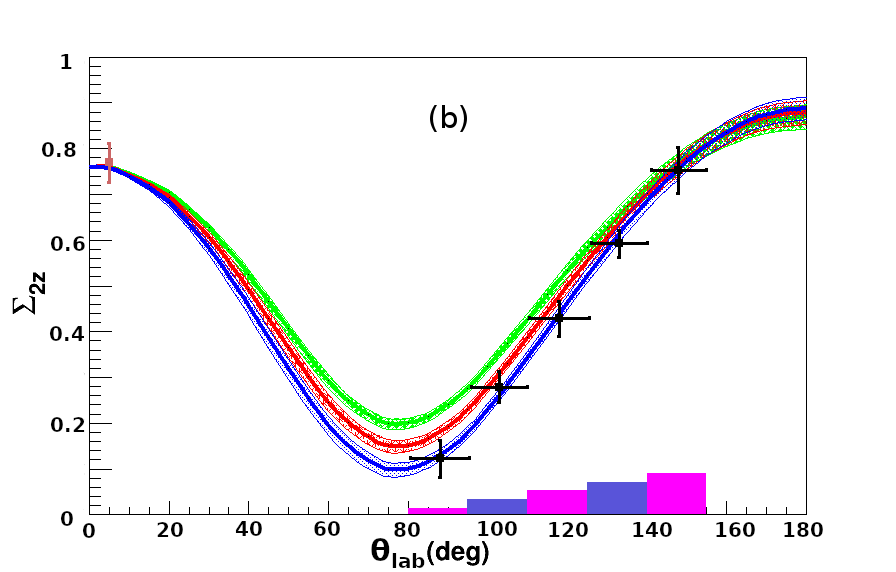}
\caption{Compton $\Sigma _{2z}$ for $E_{\gamma} = 265 - 285~$MeV (a) and
  $E_{\gamma} = 285 - 305~$MeV (b).
  The red point is the value for $\Sigma_{2z}$ at $0^{\circ}$, plotted at
  $5^{\circ}$ for readability, as determined by dispersive sum
  rules~\cite{gry15,gry16}.
  The curves are from the HDPV dispersion theory calculation of Pasquini
  {\em et al.},~\cite{holst00,drech03,pasq07}, where $\gamma_{E-}$~\cite{gri18} is fixed
  at $-3.5 \times 10^{-4}~\mathrm{fm}^{4}$ and $\gamma_{M-}$~\cite{gri18} is set
  at $-0.5$, $1.5$, or $3.5 \times 10^{-4}~\mathrm{fm}^{4}$, in the green, red,
  or blue bands, respectively.
  The width of each band represents the other parameters, $\gamma_{0}$,
  $\gamma_{\pi}$, $\alpha_{E1}+\beta_{M1}$ and $\alpha_{E1}-\beta_{M1}$ varying
  within their experimental errors.
  The error bars shown are point-to-point statistical plus random systematic
  errors added in quadrature. The correlated systematic uncertainties are shown
  as a separate block above the horizontal axis for each Compton angle. }
\label{fig-sigma2z}
\end{figure}

The $\Sigma_{2z}$ asymmetries for $E_\gamma = 265-285$~MeV and
$E_\gamma = 285-305$~MeV, obtained by combining the results from the two
beamtimes via their weighted average, are tabulated in Tab.~\ref{table:Asy} and
shown in Fig.~\ref{fig-sigma2z} along with determinations at $0^{\circ}$ through
dispersive sum rules~\cite{gry15,gry16}. While the absolute statistical errors
only vary between 0.029--0.056, the relative errors vary between 5--33\% due to
the small asymmetry at $90^{\circ}$.
The systematic errors from the three different sources: target polarization
$(10\%)$, beam polarization $(2.7\%)$, and carbon subtraction $(3-6\%)$, were
added in quadrature and their average between the 2014 and 2015 beamtimes for
each Compton angle is listed in the table and shown as a separate block above
the horizontal axis in the figures. These total systematic errors vary between
0.016--0.085 absolute, or 11--13\% relative.
To study the sensitivity of the $\Sigma_{2z}$ results on the SPs, a fixed-$t$
dispersion relation code (HDPV)~\cite{holst00,drech03,pasq07} was used to
generate predicted asymmetries at fixed lab energies for various values of the
scalar and spin polarizabilities.
Predictions within Baryon Chiral Perturbation Theory (B$\chi$PT)~\cite{pascal10}
and Heavy Baryon Chiral Perturbation Theory (HB$\chi$PT)~\cite{gri16,gri18} are
also available, but are not shown in Fig.~\ref{fig-sigma2z} to preserve readability.
The code used nominal values for the scalar polarizabilities of:
$\alpha_{E1} + \beta_{M1} = 13.8\pm0.4$ (Baldin sum rule)~\cite{leon01} and
$\alpha_{E1} - \beta_{M1} = 8.7\pm0.7$ (in units of $10^{-4}~\mathrm{fm}^{3}$)~\cite{pdg18},
and for the SPs of:
$\gamma_{0} = -0.929\pm0.105$~\cite{gry15,gry16} and $\gamma_{\pi} = 8\pm1.8$ (in units of $10^{-4}~\mathrm{fm}^{4}$)~\cite{camen02}.
It should be noted that the value for $\alpha_{E1} - \beta_{M1}$ was chosen as the current PDG numbers~\cite{pdg18}, despite the debate regarding them~\cite{grhm12,pas19}, as the focus of this study is on the spin polarizabilities.
It should also be noted that this value for $\gamma_{\pi}$ does not include the $\pi^{0}$-pole component, set as $-46.7\times10^{-4}~\mathrm{fm}^{4}$~\cite{pas19} in all of these studies.

Though $\gamma_{0}$ and $\gamma_{\pi}$ can form a basis of the SPs with
$\gamma_{E1E1}$ and $\gamma_{M1M1}$, they can alternatively form an orthogonal
basis with $\gamma_{E-}=\gamma_{E1E1}-\gamma_{E1M2}$ and
$\gamma_{M-}=\gamma_{M1M1}-\gamma_{M1E2}$~\cite{gri18}.
In Fig.~\ref{fig-sigma2z}, $\gamma_{E-}$ was fixed at
$-3.5 \times 10^{-4}~\mathrm{fm}^{4}$ and $\gamma_{M-}$ was set at $-0.5$, $1.5$,
or $3.5$ in the same units.
The various bands represent the different values for $\gamma_{M-}$, while
the spread of each band is a result of allowing $\gamma_{0}$, $\gamma_{\pi}$,
$\alpha_{E1}+\beta_{M1}$ and $\alpha_{E1}-\beta_{M1}$, to vary by their
experimental errors.
It is clear from Fig.~\ref{fig-sigma2z} that the $\Sigma_{2z}$ data in this
energy range indicate a sensitivity to $\gamma_{M-}$ of approximately $\pm2$ in
the standard units.
Alternatively, $\gamma_{M-}$ can be fixed at
$1.5 \times 10^{-4}~\mathrm{fm}^{4}$ and $\gamma_{E-}$ set at $-5.5$, $-3.5$, or
$-1.5$ in the same units.
Unlike the previous case, $\Sigma_{2z}$ in this energy range showed a weak
sensitivity to $\gamma_{E-}$.

\begin{table}[ht]
\centering
\begin{tabular}{c|rl|rl|rl}
   \hline \hline
   & \multicolumn{6}{c}{$\Sigma_{2z}$, $\Sigma_{2x}$, and $\Sigma_{3}^{LEGS}$ data fits} \\ \cline{2-7}                  
                   & \multicolumn{2}{|c|}{HDPV} & \multicolumn{2}{|c|}{B$\chi$PT} & \multicolumn{2}{|c}{Weighted average} \\ \hline
   $\gamma_{E1E1}$  & $-3.18$ & $\pm\;\,0.52$  & $-2.65$ & $\pm\;\,0.43$       &  $-2.87$ & $\pm\;\,0.52$ \\ 
   $\gamma_{M1M1}$  & $ 2.98$ & $\pm\;\,0.43$  & $ 2.43$ & $\pm\;\,0.42$       &  $ 2.70$ & $\pm\;\,0.43$ \\  
   $\gamma_{E1M2}$  & $-0.44$ & $\pm\;\,0.67$  & $-1.32$ & $\pm\;\,0.72$       &  $-0.85$ & $\pm\;\,0.72$ \\ 
   $\gamma_{M1E2}$  & $ 1.58$ & $\pm\;\,0.43$  & $ 2.47$ & $\pm\;\,0.42$       &  $ 2.04$ & $\pm\;\,0.43$ \\ \hline 
   $\chi^{2}/dof$   & 1.14    &                & 1.36    &                     &          &               \\ \hline\hline
\end{tabular}
\caption{Polarizabilities in $10^{-4}~\mathrm{fm}^{4}$ from fitting $\Sigma_{2z}$, $\Sigma_{2x}$, and $\Sigma_{3}^{LEGS}$ asymmetries using either a HDPV~\cite{holst00,drech03,pasq07} or a B$\chi$PT~\cite{pascal10} calculation, and weighted average of the SPs.}
\label{table:Sps}
\end{table}

\begin{figure}[ht]
\centering
\includegraphics[width=8.6 cm]{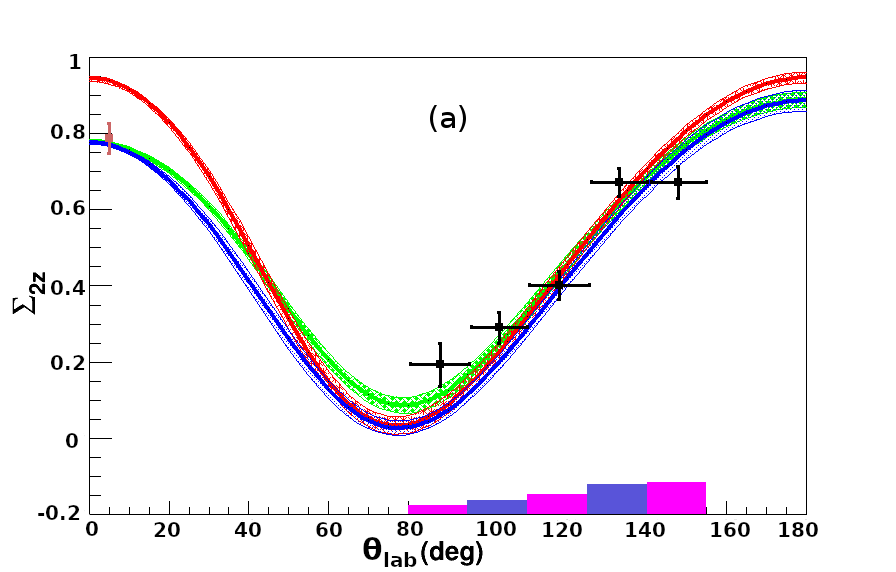}\\
\includegraphics[width=8.6 cm]{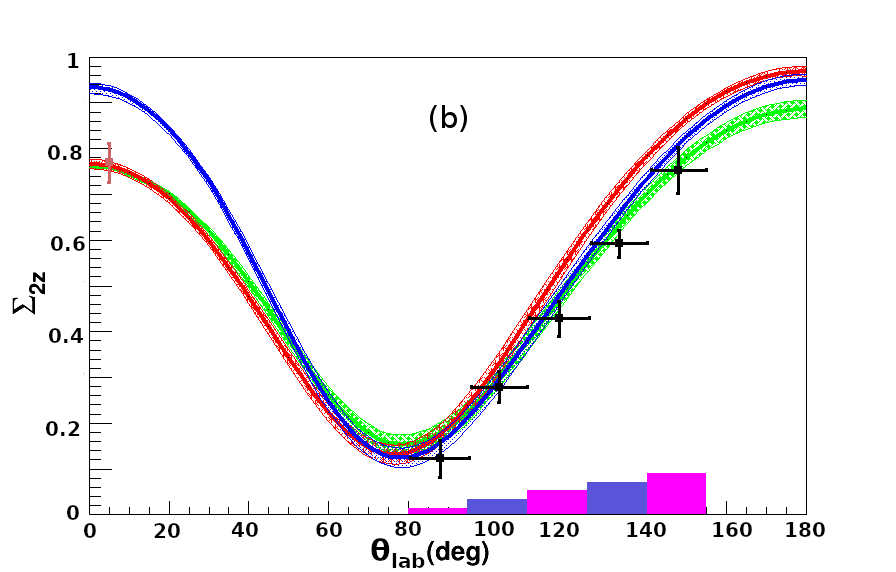}
\caption{Compton $\Sigma _{2z}$ for $E_{\gamma} = 265 - 285~$MeV (a) and
  $E_{\gamma} = 285 - 305~$MeV (b).
  The red point is the value for $\Sigma_{2z}$ at $0^{\circ}$, plotted at
  $5^{\circ}$ for readability, as determined by dispersive sum
  rules~\cite{gry15,gry16}.
  The green, red, and blue curves are from HDPV~\cite{holst00,drech03,pasq07},
  B$\chi$PT~\cite{pascal10}, and HB$\chi$PT~\cite{gri16,gri18} calculations,
  respectively. For each, the central curve uses the weighted average values from
  Table~\ref{table:Sps}, and the width of each band represents the parameters varying
  within the errors quoted in the same table.
  The error bars shown are point-to-point statistical plus random systematic
  errors added in quadrature. The correlated systematic uncertainties are shown
  as a separate block above the horizontal axis for each Compton angle. }
\label{fig-sigma2z-fit}
\end{figure}

A global analysis of $\Sigma_{2z}$ data from this measurement, along with the
published $\Sigma_{2x}$ and $\Sigma_{3}^{LEGS}$ results, and the prior values of
$\gamma_{0}$ and $\gamma_{\pi}$, was performed to study the model dependence
and extract the SPs.
This was done by fitting the asymmetry data using the
HDPV calculation~\cite{holst00,drech03,pasq07} and a B$\chi$PT calculation~\cite{pascal10}.
The extracted SPs determined using each model are summarized in Table~\ref{table:Sps}.
The fit with HDPV results in $\gamma_{E-}=-2.74 \times 10^{-4}~\mathrm{fm}^{4}$ and
$\gamma_{M-}=1.4$, in the same units, similar to the values used for the theoretical
bands in Fig.~\ref{fig-sigma2z}. The values from the two models are fairly consistent,
and the best estimate of a central value is given by the weighted average in the last
column of Table~\ref{table:Sps}. The errors for the weighted average values were
conservatively taken as the larger of the two fits.  These errors were chosen in favor
of the weighted error, because the weighted errors assume the uncertainties in the
theoretical calculations are uncorrelated, for which this paper makes no statement.
The data are again shown in Fig.~\ref{fig-sigma2z-fit}, now with theoretical
calculations for HDPV~\cite{holst00,drech03,pasq07}, B$\chi$PT~\cite{pascal10}, and
HB$\chi$PT~\cite{gri16,gri18}, using the weighted average values for the SPs.

In summary, model-dependent extractions of the SPs from a combined data fit of
double- and single-polarized Compton scattering asymmetry results in the
$\Delta(1232)$ resonance region are presented. These extracted SPs are also in
good agreement with dispersion relation~\cite{holst00,drech03,pasq07}, Baryon
Chiral Perturbation Theory~\cite{lensky15}, Heavy Baryon Chiral Perturbation
Theory~\cite{gri16,gri18}, K-matrix theory~\cite{kondra01}, and chiral
Lagrangian~\cite{gas11} predictions.
Although the uncertainties in the SPs are significantly improved compared to
previously reported results~\cite{martel15}, forthcoming $\Sigma_{3}$ results
from MAMI experiments~\cite{col17} are expected to provide further improvements
in the determination of these fundamental nuclear structure terms. 

\begin{acknowledgments}
  The authors wish to acknowledge the excellent support of the accelerator group and operators of MAMI\@.
  We also wish to acknowledge and thank B. Pasquini and V. Pascalutsa for the use of their theory codes, as well as H. Grie{\ss}hammer, J. McGovern, D. Phillips, and M. Vanderhaeghen for their theoretical support.
  This work was supported by the Natural Sciences and Engineering Research Council of Canada (NSERC, FRN: SAPPJ-2015-00023),
  the U.S. Department of Energy (Offices of Science and Nuclear Physics, Award Nos.\ DE-FG02-99-ER41110, DE-FG02-88ER40415, DEFG02-01-ER41194, DE-SC0014323)
  and National Science Foundation (Grant Nos.\ PHY-1039130, PHY-1714833, OISE-1358175),
  Deutsche Forschungsgemeinschaft (SFB443, SFB/TR16, and SFB1044),
  the European Community-Research Infrastructure Activity under the FP6 ``Structuring the European Research Area'' program (Hadron Physics, Contract No.\ RII3-CT-2004-506078),
  Schweizerischer Nationalfonds (Contract Nos.\ 200020-156983, 132799, 121781, 117601, and 113511),
  the U.K. Science and Technology Facilities Council (STFC 57071/1, 50727/1, ST/L00478X/1, and ST/L005824/1),
  and INFN (Italy).
\end{acknowledgments}

\end{document}